# Quantum Measurement Complexity

Subhash Kak

**Abstract.** This paper explores the problem of quantum measurement complexity. In computability theory, the complexity of a problem is determined by how long it takes an effective algorithm to solve it. This complexity may be compared to the difficulty for a hypothetical oracle machine, the output of which may be verified by a computable function but cannot be simulated on a physical machine. We define a quantum oracle machine for measurements as one that can determine the state by examining a single copy. The complexity of measurement for a realizable machine will then be respect to the number of copies of the state that needs to be examined. A quantum oracle cannot perform simultaneous exact measurement of conjugate variables, although approximate measurement may be performed as circumscribed by the Heisenberg uncertainty relations. When considering the measurement of a variable, there might be residual uncertainty if the number of copies of the variable is limited. Specifically, we examine the quantum measurement complexity of linear polarization of photons that is used in several quantum cryptography schemes and we present a relation using information theoretic arguments. The idea of quantum measurement complexity is likely to find uses in measurements in biological systems.

**Introduction**

In computer science, an oracle machine is an abstract machine for decision problems [1]. Like the oracle of Greek mythology, it is able to decide a decision problem in a single operation. It may be visualized as a black box the practicality of whose implementation is of no concern. There are two kinds of oracle: the exact one, and the one that gives its response in a probabilistic manner.

The computational problem that is solved in one step by an oracle may require several steps to solve if an algorithm is used. If the computational effort to solve the problem increases as a polynomial in the size of the problem, the problem is said to belong to the class P. Another way to look at an oracle is as a nondeterministic polynomial algorithm that solves the decision problem in a single step. Problems solved by such an oracle based procedure are said to belong to the class NP; thus the class P is the set of problems that can be solved in polynomial time by an algorithm without the use of an oracle. One of the fundamental unsolved problems of computer science is whether P=NP.

There exists a theory of quantum computational complexity [2] which deals with the class of decision problems solvable by a quantum computer. For example the class BQP represents decision problems solvable by a quantum computer in polynomial time with an error probability that is at most 1/3. The theory of quantum computational complexity deals with the circuit model of quantum computation (for issues of implementation of the circuit model [3]-[5]) that does not concern us here. Our concern is with quantum measurement and not computation.

The process of measurement is subject to noise, measurement errors, and other kinds of uncertainty. When considering the quantum mechanical framework one usually speaks of the uncertainty in the simultaneous accuracy in the measurement of conjugate variables and the randomness associated with the collapse of the wavefunction upon measurement [6]. There is additional uncertainty that needs to be taken into consideration when accounting for interaction of the system with the environment and this is generally labeled decoherence or quantum noise. In certain formulations, the collapse of the wavefunction is viewed as an outcome of the interaction between the primary quantum system and the environment, which is viewed as another quantum system [7].

An *effective measurement* may be defined as a value that is correct to within prespecified uncertainty. We can use the perspective of the oracle machine to examine this problem. Given the measurement of one of the conjugate variables, not even an oracle can determine the other variable exactly since a quantum state cannot be simultaneous eigenstate of both the variables. If the variables are not conjugate, the oracle can provide us values of both. On the other hand, there can be non-conjugate variables, where knowledge of one of the variables is essential in measurement of the other variable if only to a certain degree of accuracy.

In the orthodox Copenhagen interpretation (CI) of quantum mechanics [6],[8], the wavefunction represents the observer's knowledge of the system. This epistemic view is to be contrasted from the ontic view [9] that the wavefunction is the reality itself. The ontic view, which is often put under the rubric of the many worlds interpretation (MWI), cannot explain why in a universe characterized by its own wavefunction, different subsystems arise that cause increase in information by the process of mutual interaction. In CI, the observer is viewed as apart from the quantum system, and therefore measurement represents the action of the sentient agent [10] and thus the quantum mechanical postulate about the collapse of the wavefunction is somehow related to the nature of the cognitive process.

Since the conclusion of the measurement process is in terms of classical variables, quantum theory itself has within it the bridge to tie it to classical mechanics. Furthermore, when cognitive processes are brought into the picture, the question on limits on observability and computability arise [11]-[13]. There is additionally the question of the philosophical significance of the process of observation on the evolution of the system as in the Quantum Zeno effect [14].

If one were to push the line of reasoning associated with the existence of conscious agents to its logical conclusion, then one might wonder about the relationship the indubitable nonlocality of quantum phenomena, as represented for example by entanglement, has with the fact that the cognitive process concludes with a definite value. It was recently proposed [15] that the classical cognitive processes place a veil on the nonlocality of the quantum reality. The question that arises then is whether such veiling also leads to uncertainty, and if so whether it is of different form than the types mentioned earlier in this section. This paper does not consider these philosophical issues related to uncertainty.

Here we are concerned with quantum measurement complexity in the limited context of polarization of photons. The polarization and changes in polarization of a propagating electromagnetic wave are described by the Poincaré sphere where each polarization state corresponds to a unique point on the sphere. The polarization properties of light are also described by the classical Stokes parameters whereas quantum Stokes parameters provide operator representations of the polarization for use in quantum optics. The operators satisfy commutation relations and the variances of the Stokes parameters are governed by uncertainty relations [16],[17]. But polarization is not in a conjugate relationship with the count of photons, therefore polarization should be determinable by an oracle.

The points on the equator of the sphere indicate linear polarizations and the two poles represent left and right-hand circularly polarized light; all other points on the sphere represent elliptical polarization states. An arbitrarily chosen point H on the equator designates horizontal linear polarization, and the diametrically opposite point V designates vertical linear polarization. In quantum cryptography schemes, the number of photons and polarization are critical variables [18],[19]. The original description of the BB84 protocol was based on the use of polarization although that is not used in current implementations [20].

In many laboratory settings heralded photons are used to generate single photons [21],[22]. These photons are generated using parametric down-conversion. But these photons come with uncertainty related to time of arrival and the actual polarization value until such time the heralded photon is measured to determine the polarization of its twin. Furthermore, the most efficient implementations provide a measurement only 84% of the time [23]. Therefore, the joint uncertainty related to both the entangled particles is much higher than indicated by the uncertainty limits being considered here.

The complexity of measurement in polarization-based cryptography applications is an important measure of security [24]-[26]. In this paper we derive a number-polarization uncertainty relation based on information theoretic considerations. If we adopt the positivist interpretation that in most situations the wavefunction represents the knowledge of the state rather than some existing reality, the derived relation may be of importance in other quantum information situations.

**Oracle machines**
The oracle machine, conceived of as a "black box" that produces a solution to a given computational problem, may be viewed as solving either a decision problem or a function mapping. The oracle is a set X, $X \subseteq \Sigma^*$ where $\Sigma^*$ is the set of all finite bit strings. When fed a string $x \in \Sigma^*$ (the query), the oracle returns a '1' if $x \in X$ and '0' otherwise. Let the function problem be represented by a function *f* from natural numbers (or strings) to natural numbers (or strings). An instance of the problem is an input *x* for *f* and the solution produced by the oracle is the value *f(x)*.

For a quantum query, *x*, let *y* be the single bit answer to the query whether x belongs to a certain set X. In other words, f(x) = 1 if x ∈ X and f(x) = 0 otherwise. The quantum oracle [27] is a unitary transformation, *U*, so that $U|x,0\rangle \to e^{i\phi_{x,0}}|x,f(x)\rangle$. Owing to unitarity, it follows that $U|x,1\rangle \to e^{i\phi_{x,1}}|x,f(x)+1\mod 2\rangle$. Quantum oracles that perform function mapping can be similarly visualized.

We can speak of two types of quantum measurement oracles depending on whether it outputs a precise number or if it is probabilistic, that is it provides the response in terms of a probability distribution. The oracle that outputs the precise value can be considered to be a special case of the probabilistic oracle where the standard deviation is zero. Clearly the measurement power (in terms of the capacity to obtain information) of the precise and the probabilistic oracles will be different. The exact value oracle cannot deal with conjugate variables, whereas the probabilistic oracle can so long as the measurements are governed by the uncertainty relations.

One may look at the measurement process as a game between the preparer of quantum states in the laboratory and the oracle or the recipient machine to whom the quantum states are transmitted. For the problem of the measurement of polarization the variables of interest are photon number and polarization. The purpose of the game is to use as few photons as is necessary to determine the state.

When a quantum oracle is used provide the value of polarization, this value can be verified by a machine. In principle, only two copies of the state are required: one for the oracle, and the other for the verifying machine. In practice, the verifying machine will use *m* copies to confirm with probability $1-2^{-m}$ that the decision of the oracle was correct. Note that the standard measuring instruments such as ones used by the verifying machine cannot determine the polarization in a single step.

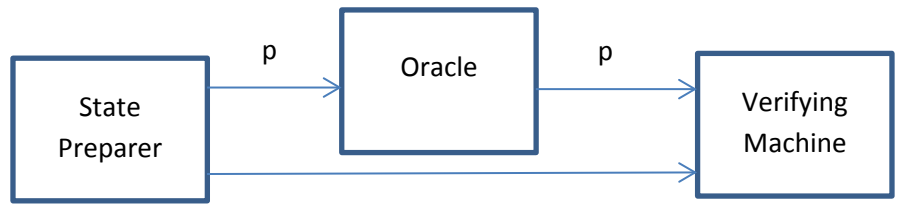

*Figure 1.* One-step oracle decision and one-step verification (in principle only two copies of the state are required)

The quantum oracle that is able to recognize an unknown quantum state is equivalent to a quantum cloner (Figure 2). If the oracle can make one copy of the unknown state, it can make any number of copies and these can then be analyzed by the use of tomography. Although, perfect cloning is forbidden by quantum mechanics, approximate cloning is possible as was shown theoretically by Bužek and Hillery [28] and implemented thereafter. In one proposal a single input photon stimulates the emission of additional photons from a source based on parametric down-conversion leading to the production of quantum clones with near optimal fidelity [29],[30]. Universal quantum cloning machines provide fidelity of $F(\rho,\sigma) \equiv tr\sqrt{\rho^{1/2}\sigma\rho^{1/2}} = \frac{5}{6}$ for two copies.

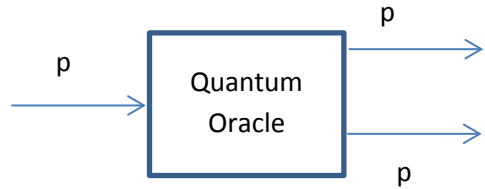

*Figure 2.* Quantum oracle as a cloning machine

A state estimator machine, using quantum state tomography, may be used to determine the value of the unknown polarization [31],[32]. If $\rho$ is the density matrix and X, Y, Z are the Pauli operators, the single qubit density matrix may be expanded as

$$\rho = \frac{tr(\rho)I + tr(X\rho)X + tr(Y\rho)Y + tr(Z\rho)Z}{2}$$

and this used to estimate the density matrix.

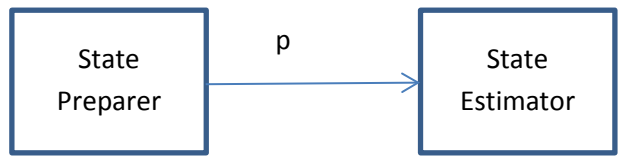

*Figure 3.* Multi-step state tomography that requires many copies of the state

The standard deviation of each of these terms for a total of $m$ measurements each is $1/\sqrt{m}$. Basically, the shape of the quantum object is pictured in the phase space using the Wigner representation. The Wigner function can be reconstructed from a set of quadrature distributions, $pr(x,\theta)$, that are projections obtained using homodyne detection, by the inverse Radon transform:

$$W(q,p) = -\frac{\wp}{2\pi^2} \int_0^{2\pi} \int_{-\infty}^{+\infty} \frac{pr(x,\theta)dxd\theta}{(q\cos\theta + p\sin\theta - x)^2}$$

where $\wp$ represents the Cauchy principal value of the integral. The homodyne detection must be performed several times on identical photons to gain sufficient statistical information. Such a reconstruction of the Wigner function can provide statistical estimates of both the conjugate variables [33].

In general, the complexity of a state estimator will be some function $f(1/n)$. As $n \to \infty$, the estimator becomes equivalent to the quantum oracle.

It is clear that access to a quantum oracle machine makes it possible to solve problems that quantum computers cannot. For example, a quantum oracle can compute the quantum Fourier transform which standard quantum circuits can only exploit indirectly [34].

If quantum oracles existed current quantum cryptography protocols will fail. Given that the existence of a quantum oracle is equivalent to the use of many photons in the same state and given the fact that perfect sources that produce single photons do not exist, the case for quantum cryptography makes muddled and one needs to go beyond old algorithms. On the other hand, if the standard BB84 protocol was used, one could simply transmit the 0s and 1s in conjugate bases and they would be received correctly by the receiver assuming there was no interference by the eavesdropper and no noise. The fact of such a correct reception could be confirmed by the exchange of hash computed by both the parties [35]. The conditions necessary to guarantee security would likewise change for the three-stage quantum cryptography protocol.

**Number-phase uncertainty and complexity**

In many quantum computing and cryptography situations it is implicitly assumed that single particles are being used. But the localization of single particles with known wavefunction is a challenge and one may speak of a complexity function associated with the number of copies of the particles that are used in practice. According to the Heisenberg uncertainty principle, for conjugate variables A and B

$$\Delta A \Delta B \geq \frac{1}{2}|\langle [A,B] \rangle|$$

where $(\Delta A)^2 = \langle A^2 \rangle - \langle A \rangle^2$. In the case of position, $x$, and momentum, $p$, the canonical commutation relation $[x,p] = i\hbar$ implies that $\Delta x \Delta p \geq \frac{\hbar}{2}$.

For a harmonic oscillator with the Hamiltonian $H = \frac{p^2}{2m} + \frac{1}{2}m\omega^2 x^2$, the annihilation operator is $a = \sqrt{\dfrac{m\omega}{2\hbar}}\left(x + \dfrac{ip}{m\omega}\right)$. The phase and the number operators, θ and N respectively, are related by

$a = e^{i\theta}\sqrt{N}$. The phase eigenstate is $|\theta\rangle = \sum_{n=1}^{\infty} e^{in\theta}|n\rangle$. This eventually leads to the number-phase uncertainty relation:

$$\Delta N \Delta \theta \geq \dfrac{1}{2}$$

We cannot speak of the value of the phase unless many copies of the object are available. A measurement scheme that allows one to reach the equality would be optimal and one would expect it to be associated with the highest level of complexity.

The number-phase relation is of significance in quantum cryptography schemes that use phase coding although in many such schemes it is not the phase but rather the difference in phase that carries the encrypted information. Certain quantum cryptography schemes require only single photons to be sent that brings about implementation issues owing to the time-energy uncertainty relations [36]. Since weak light sources produce stream of photons that are given by the Poisson distribution, the presence of a single photon in the time window of interest can only be guaranteed probabilistically. This is also related to the initialization of the quantum state in the quantum information scheme [37],[38].

**Number-polarization complexity**
The consideration of number polarization uncertainty is motivated by questions regarding the nature of the photon and the amount of information it can carry. Photons carry both spin and orbital angular momentum (OAM). The spin is associated with polarization and the OAM with the azimuthal phase of the complex electric field. A photon with azimuthal phase dependence of the form $e^{il\phi}$ carries OAM equal to $l\hbar$. Whereas a photon can at most carry one bit of information in its polarization, the infinite number of orthogonal states of OAM means that there is no limit on the number of bits that can be carried by a single photon [39],[40].

In quantum fields polarization is determined in terms of the mean values of the Stokes operators which satisfy spin algebra. One can, thereupon, use tomography on the Stokes operators and measure uncertainty in the polarization value related to the number of photons.

Here we consider the special case of coherent photons that are linearly polarized as in the aforementioned cryptography applications. The Stokes operators satisfy the commutation relations:

$$[S_i, S_j] = 2i\varepsilon_{ijk} S_k, \qquad i, j, k = 1, 2, 3$$

Let the given photons be in an unknown coherent state:

$$|\varphi\rangle = \alpha|H\rangle + \beta|V\rangle$$

where $|H\rangle$ represents a horizontally polarized state and $|V\rangle$ represents a vertically polarized state. The Stokes parameters are then simply:

$$\langle S_0 \rangle = \alpha^*\alpha + \beta^*\beta \; ; \; \langle S_1 \rangle = \alpha^*\alpha - \beta^*\beta$$
$$\langle S_2 \rangle = \alpha^*\beta + \alpha\beta^* \; ; \; \langle S_3 \rangle = -i(\alpha^*\beta - \alpha\beta^*)$$

Note that $S_0$ commutes with the other $S_i$. Since $S_0$ represents the number of photons, this implies that the knowledge of this number should not have an impact in our determination of the polarization of the photons.

Now we turn the problem around and consider polarization from the perspective of measurement or tomography. Given an unknown state, it can obviously not be determined given a single copy of the state. Let the accuracy of the polarization angle of the photons is Δk which may be seen as the size of a bin on the equator of the Poincaré sphere.

We need to consider only half the equator for say horizontal polarization, that is a total of $\pi$, since the rest of the equator will be defined by the vertical polarization. Over this range we can choose $2^m$ bins of accuracy. The accuracy of Δk, which represents the angle of each bin, corresponds to a total of $\frac{\pi}{\Delta k} = 2^m$ bins on the half-equator of the Poincaré sphere.

We assume that the photons can be in any of the $\frac{\pi}{\Delta k} = 2^m$ bins with equal probability, which represents the case that corresponds to the most uncertainty. These m bins may be compared to the characteristics of a probabilistic oracle. Since the variables are not conjugate we need not consider least-variance distributions such as the von Mises distribution on the circle [41].

From an information theoretic point of view, we need m photons to distinguish between $2^m$ possible states using a divide-and-conquer strategy. Therefore, to obtain the stated accuracy of Δp, the number of photons, *m*, should be taken to be ΔN. This means that

$$\Delta N \Delta k \approx m \frac{\pi}{2^m}$$

When m=1, the uncertainty relation is:

$$\Delta N \Delta k \approx \frac{\pi}{2}$$

As $m \to \infty$, the uncertainty goes to zero.

It is possible that the above heuristic considerations over-estimated the uncertainty and the actual result should be:

$$\Delta N \Delta k \geq \frac{1}{2}$$

If not, this would be another differentiator between classical and quantum states in the style of Bell inequalities [42],[43].

An interesting problem to investigate is to determine the complexity of operations associated with the measurement of polarization computed with respect to both the number of copies examined and the computations performed to compute the result.

**Measurement complexity in biological systems**
The idea of quantum oracle can also be helpful in studying the measurement complexity of biological systems in which quantum mechanics is increasingly seen to play a direct or indirect role. Quantum mechanics has been invoked to explain photosynthesis [44], olfaction [45], vision [46], long-range electron transfer [47], and bird navigation [48], and memory [49],[50]. For example, quantum coherence persists in photosynthesis in spite of the molecular noise present in the cells. Dissipative structures can support quantum effects associated with virtual particles. Macromolecule vibrations create quasiparticles and therefore quantum effects associated with such quasiparticles are contingent on specific macro-structures [51],[52].

Further studies of these examples of quantum processing [53] will need to go from understanding of the basic processes to determining is such processing can also be put in different classes in terms of measurement complexity.

It is plausible that the quantumness of certain variables in the biological domain has not been identified due to the high complexity of the measurement process.

**Conclusions**

This paper explores the idea of quantum measurement complexity. It began by speaking of oracles in the context of effective measurement of a variable based on the number of copies that are being considered. To consider the ideal case, we defined the quantum measurement oracle in analogy with the classical oracle in that it can estimate an unknown state.

Simultaneous exact measurement of conjugate variables cannot be performed by a quantum oracle, although approximate measurement may be performed. A quantum oracle makes it possible to perform cloning which means that a quantum machine that uses oracles is equivalent to the use of several photons rather than a single one for each qubit.

In the actual measurement of variables that are not conjugate, one may speak of uncertainty associated with the measurement of unknown value. Linear polarization of photons is used in several quantum cryptography schemes and, therefore, an uncertainty relation concerning polarization is of interest. A more refined number-polarization uncertainty relation, together with a corresponding algorithm to obtain optimal estimate, can help determine bounds on the number of photons that can be used in multi-photon quantum cryptography to guarantee a certain level of security. Such a study can also include tomography of hyper-entangled photons [54]. Another interesting problem is to investigate approximate quantum cloning in examining bounds on the photon number polarization relation.

We expect a further examination of quantum measurement complexity to have particular value in the study of biological systems.

**References**
[1] Baker, T. P., Gill, J., Solovay, R. Relativizations of the P = NP question. SIAM Journal on Computing 4 (4): 431–442 (1975)
[2] Watrous, J. Quantum computational complexity. Encyclopedia of Complexity and Systems Science. Springer (2009)
[3] Kak, S. General qubit errors cannot be corrected. Information Sciences 152: 195-202 (2003)
[4] Dyakonov, M. Revisiting the hopes for scalable quantum computation. arXiv:1210.1782
[5] Kak, S. The problem of testing a quantum gate. Infocommunications Journal 4: 18-22 (2012)
[6] Heisenberg, W. Physics and Philosophy. London: George Allen & Unwin (1958)
[7] Zurek, W.H. Decoherence, einselection, and the quantum origins of the classical. Rev. Mod. Phys. 75: 715-775 (2003)
[8] Von Neumann, J. Mathematical Foundations of Quantum Mechanics. Princeton, NJ: Princeton University Press (1932/1955)
[9] Tegmark, M. (1998) The interpretation of quantum mechanics: many worlds or many words? Fortsch. Phys. 46: 855-862.
[10] Stapp, H.P. Mind, Matter, and Quantum Mechanics. New York: Springer-Verlag (2003)
[11] Kak, S. The Nature of Physical Reality. New York: Peter Lang (1986/2011)
[12] Kak, S. Active agents, intelligence, and quantum computing. Information Sciences 128: 1-17 (2000)

[13] Kak, S. Observability and computability in physics, Quantum Matter 3: 172-176 (2014)
[14] Misra, B. and Sudarshan, E. C. G. The Zeno's paradox in quantum theory. Journal of Mathematical Physics 18: 758–763 (1977)
[15] Kak, S. From the no-signaling theorem to veiled non-locality. NeuroQuantology 12: 12-20 (2014)
[16] Agarwal, G.S. Quantum Optics. Cambridge University Press (2013)
[17] Klauder, J.R. and Sudarshan, E.C.G. Fundamentals of Quantum Optics. Dover (2006)
[18] Bennett, C.H. and Brassard, G. Quantum cryptography: Public key distribution and coin tossing. Proceeding of the IEEE International Conference on Computers, Systems, and Signal Processing, Bangalore, India, pp. 175–179, IEEE, New York (1984)
[19] Kak, S. A three-stage quantum cryptography protocol. Foundations of Physics Letters 19: 293-296 (2006)
[20] Gerhardt, I., Liu, Q., Lamas-Linares, A., Skaar, J., Kurtsiefer, C., and Makarov, V. Full-field implementation of a perfect eavesdropper on a quantum cryptography system. Nat. Commun. 2, 349 (2011)
[21] Kwiat, P.G., Mattle, K., Weinfurter, H., Zeilinger, A., Sergienko, A.V., and Shih, Y. New high-intensity source of polarization-entangled photon pairs. Phys. Rev. Lett. 75, 4337–4341 (1995).
[22] Pittman, T., Jacobs, B., and Franson, J. Heralding single photons from pulsed parametric down-conversion. Opt. Commun. 246: 545–550 (2005)
[23] Cunha, D. et al. Demonstrating highly symmetric single-mode, single-photon heralding efficiency in spontaneous parametric downconversion. Optics Letters, 38, 1609 (2013)
[24] Kak, S. Quantum information and entropy. International Journal of Theoretical Physics 46: 860-876 (2007)
[25] Chen, Y. et al. Multi-photon tolerant secure quantum communication -- from theory to practice. Proceedings International Communications Conference (ICC), Budapest (2013)
[26] Kak, S. Threshold quantum cryptography. arXiv:1310.6333
[27] Deutsch, D. and Jozsa, R. Rapid solution of problems by quantum computation. Proc. R. Soc. Lond. A, 439:553 (1992)
[28] Bužek V. and Hillery, M. Quantum copying: beyond the no-cloning theorem. Phys. Rev. A 54, 1844 (1996)
[29] Lamas-Linares, A., Simon, C., Howell, J.C., Bouwmeester, D. Experimental quantum cloning of single photons. Science 296, 712 (2002)
[30] Scarani, V., Iblisdir, S., Gisin, N. Quantum cloning. Rev. Mod. Phys. 77: 1225-1256 (2005)
[31] Chiribella, G. and Yang, Y. Optimal asymptotic cloning machines. arXiv:1404:0990
[32] Leonhardt, U. Measuring the Quantum State of Light. Cambridge University Press (2005)
[33] Altepeter, J.B., Jeffrey, E.R., and Kwiat, P.G. Photonic state tomography. In Advances in Atomic, Molecular, and Optical Physics. Elsevier (2005)
[34] Nielsen, M.I. and Chuang, I.L. Quantum Computation and Quantum Information. Cambridge University Press, 2000.
[35] Kak, S. The piggy bank cryptographic trope. Infocommunications Journal 6: 22-25 (2014)
[36] Hilgevoord, J. The uncertainty principle for energy and time I. American Journal of Physics 64, 1451-1456 (1996)
[37] Kak, S. The initialization problem in quantum computing. Foundations of Physics 29: 267-279 (1999)
[38] Bruno, A., Capolupo, A., Kak, S., Raimondo, G., and Vitielli, G. Gauge theory and two level systems. Mod. Phys. Lett. B 25: 1661-1670 (2011)

[39] Molina-Terriza, G., Torres, J.P., Toerner, L. Twisted photons. Nature physics 3: 305-310 (2007)
[40] Leach, J. et al., Measuring the orbital angular momentum of a single photon. Phys. Rev. Lett. 88: 257901 (2002)
[41] Best, D. and Fisher, N. Efficient simulation of the von Mises distribution. Applied Statistics, 28, 152–157 (1979)
[42] Bell, J. On the Einstein Podolsky Rosen paradox. Physics 1 (3): 195–200 (1964)
[43] Kak, S. Probability constraints and the classical/quantum divide. NeuroQuantology 11: 600-606 (2013)
[44] Collini, E. et al. Coherently wired light-harvesting in photosynthetic marine algae at ambient temperature. Nature 463, 644–648 (2010)
[45] Turin, L. A spectroscopic mechanism for primary olfactory reception. Chem. Senses 21, 773–791 (1996)
[46] Polli, D. et al. Conical intersection dynamics of the primary photoisomerization event in vision. Nature 467, 440–443 (2010)
[47] Gray, H. B. & Winkler, J. R. Long-range electron transfer. Proc. Natl Acad. Sci. USA 102, 3534–3539 (2005).
[48] Ritz, T., Thalau, P., Phillips, J. B., Wiltschko, R. & Wiltschko, W. Resonance effects indicate a radical pair mechanism for avian magnetic compass. Nature 429, 177–180 (2004)
[49] Kak, S. Artificial and biological intelligence. ACM Ubiquity 6 (42): 1-22 (2005)
[50] Kak, S. Biological memories and agents as quantum collectives. NeuroQuantology 11: 391-398 (2013)
[51] Freeman, W. and Vitiello G, Nonlinear brain dynamics as macroscopic manifestation of underlying many-body dynamics. Physics of Life Reviews 3: 93-118 (2006)
[52] Kak, S. Quantum neural computing. In Advances in Imaging and Electron Physics, vol. 94, P. Hawkes (editor). Academic Press, 259-313 (1995)
[53] Lambert, N. et al, Quantum biology. Nature Physics 9, 10–18 (2013)
[54] Barreiro, J.T., Langford, N.K., Peters, N.A., Kwiat, P.G. Generation of hyperentangled photon pairs. Phys. Rev. Lett. 95: 260501 (2005)